\documentclass[twocolumn,showpacs,preprintnumbers,amsmath,amssymb,floatfix]{revtex4}

\usepackage{graphicx}
\usepackage{dcolumn}
\usepackage{bm}
\usepackage{epsfig}

\begin{document}

\preprint{APS/123-QED}

\title{Predictions of bond percolation thresholds for the kagom\'e
and Archimedean (3,12$^2$) lattices}
\author{Christian R. Scullard}
\email{scullard@uchicago.edu}
\affiliation{Department of Geophysical Sciences, University of Chicago, Chicago, Illinois 60637, USA}
\author{Robert M. Ziff}
\email{rziff@umich.edu}
\affiliation{Michigan Center for Theoretical Physics and Department of Chemical Engineering, University of Michigan, Ann Arbor, Michigan 48109-2136, USA}

\date{6 May 2006}

\begin{abstract}
Here we show how the recent exact determination of the bond percolation threshold for the martini lattice can be used to provide approximations to the unsolved kagom\'e and $(3,12^2)$ lattices. We present two different methods, one of which provides an approximation to the inhomogeneous kagom\'e and $(3,12^2)$ bond problems, and the other gives estimates of $p_c$ for the homogeneous kagom\'e ($0.5244088...$) and $(3,12^2)$ ($0.7404212...$) problems that respectively agree with numerical results to five and six significant figures.
\end{abstract}

\maketitle

Percolation \cite{Grimmett, Stauffer} has provided some of the most intriguing and difficult problems in statistical mechanics. Devised in 1957 by Broadbent and Hammersley \cite{BroadbentHammersley}, it has served as the simplest example of a lattice process exhibiting a phase transition, and its study provides insight into more complicated physical models.

\begin{figure}
\begin{center}
\includegraphics[]{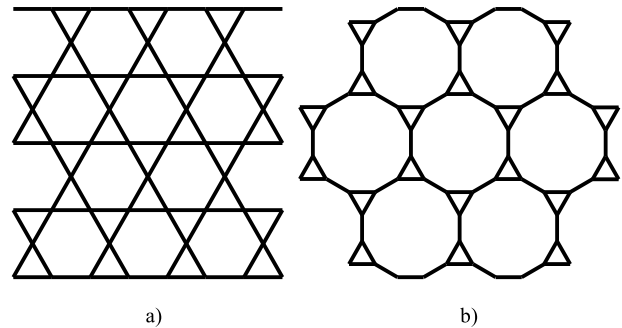}
\caption{(a) the kagom\'e lattice, (b) the $(3,12^2)$ (or 3-12) lattice.} \label{fig:kagome_3_12}
\end{center}
\end{figure}

The problem is very simply stated. Given any lattice, such as either of those shown in Fig.\ \ref{fig:kagome_3_12}, we declare each bond to be in one of two states; open or closed. If a bond (although we could just as well consider sites) is open with probability $p$ and closed with probability $1-p$, then clusters of various sizes will appear, with the average cluster size increasing as a function of $p$. In the limit of an infinite lattice there exists a critical value of this parameter, denoted $p_c$ and referred to as the percolation or critical threshold, where an infinite cluster will appear with probability $1$. The value of $p_c$ is specific to each lattice.

While the problem can be easily and precisely defined, exact solutions for thresholds (or anything else for that matter) have historically proved elusive, with results being limited to a small set of lattices. Recent work \cite{Ziff06,Scullard06} has significantly expanded this set, and in fact it was shown in \cite{Ziff06} that an infinite variety of problems are exactly solvable so long as their basic cells are contained between three vertices and are stacked in a particular self-dual way. Despite this recent progress, the most perplexing unsolved problems still remain. In particular, the exact site percolation thresholds of the square and honeycomb (also called hexagonal) lattices, and the bond threshold of the kagom\'e lattice are still unknown after nearly half a century of research in the field. The latter problem is one of the subjects of this Communication.

The square, honeycomb, and kagom\'e problems belong to an important subset of two dimensional lattices called the Archimedean lattices \cite{Grunbaum}, in which all sites are equivalent. There are 11 such graphs, and although both site \cite{SudingZiff99} and bond \cite{Parviainen} thresholds have been studied numerically for all of them, the only exactly solved problems are the bond thresholds of the square, honeycomb, and triangular \cite{SykesEssam} lattices, and the site thresholds of the triangular, kagom\'e, and $(3,12^2)$ lattices. Note that finding the site threshold is a completely different problem from finding the bond threshold, and these last two site values are known only because of a trivial transformation from the honeycomb bond lattice --- a transformation that does not help us in solving the bond problems. However, the $(3,12^2)$ lattice bears enough similarity to the kagom\'e that the methods we present here will provide us with estimates for that bond threshold as well, one of which agrees with a recent numerical result \cite{Parviainen} to its limit of precision, which is six significant figures.

The bond threshold for the kagom\'e lattice has previously been the subject of several conjectures \cite{Wu06,Tsallis,Chao,HoriKitahara}. Using a method that predicted correct critical frontiers for the Potts model \cite{Wu82} on other lattices, Wu \cite{Wu79} conjectured that it would also work for the kagom\'e, and, using the fact that percolation is the $q \rightarrow 1$ limit of the Potts model \cite{FortuinKasteleyn}, proposed that $p_c=0.524430...$, the solution of a polynomial we will encounter below. A few years afterward, and also in the context of the Potts model, Tsallis \cite{Tsallis, Tsallis2} offered the competing conjecture $p_c=0.522372...$, employing an argument that also made correct predictions for other lattices. It was not until much later that both of these propositions were ruled out numerically \cite{ZiffSuding97} though fairly high precision was required to exclude Wu's estimate. Tsallis also considered the $(3,12^2)$ lattice, and proposed $p_c(3,12^2)=0.739830...$\ .

Aside from these various speculative methods, in which one makes conjectures that must be verified or rejected numerically, there are some rigorous results for the kagom\'e and $(3,12^2)$ thresholds in the form of bounds on the values of $p_c$. This work is largely carried out by Wierman and co-workers \cite{Wierman03, MayWierman05}, using a technique called substitution. The method is such that continual refinements are possible and the most current rigorous bounds are \cite{MayWierman06}:
\begin{equation}
0.522197<p_c(\hbox{kagom\'e})<0.526873 \ ,
\end{equation}
and
\begin{equation}
0.739773<p_c(3,12^2)<0.741125 \ . \label{eq:may_wierman}
\end{equation}

Various other quantities besides the standard percolation threshold have also been studied on the kagom\'e lattice such as the mixed site-bond threshold \cite{Tarasevich}, a correlated percolation threshold \cite{Mendelson}, and an exact solution for the average cluster number on a kagom\'e lattice strip \cite{ChangShrock}, among others. As already mentioned, the kagom\'e Potts model has also received, and continues to receive, attention. In addition to the work already cited, some recent examples include \cite{Monroe}, and \cite{Jensen} in which the conjectures of Wu and Tsallis are discussed for various values of $q$.

\begin{figure}
\begin{center}
\includegraphics[]{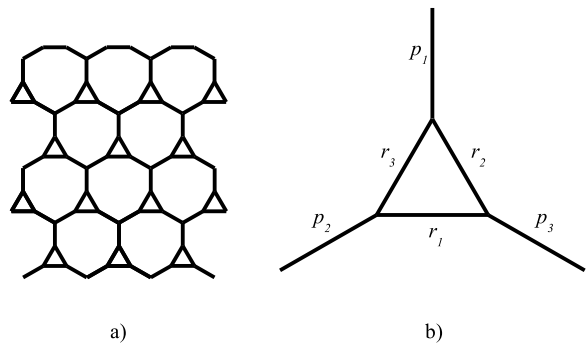}
\caption{a) The martini lattice, b) The assignment of probabilities for the inhomogeneous threshold } \label{fig:martini}
\end{center}
\end{figure}

Here we show how a recent exact solution on a similar lattice, the martini lattice [Fig. \ref{fig:martini}(a)], can be used to provide precise
estimates of the kagom\'e and $(3,12^2)$ thresholds.

The starting point of our analysis is the bond threshold for the martini lattice [Fig.\ \ref{fig:martini}(a)]. For the general martini generator of Fig.\ \ref{fig:martini}(b), the method outlined in \cite{Ziff06} gives for the inhomogeneous critical surface
\begin{eqnarray}
1 &-& p_1 p_2 r_3 - p_2 p_3 r_1 - p_1 p_3 r_2 - p_1 p_2 r_1 r_2 \nonumber \\
&-& p_1 p_3 r_1 r_3 - p_2 p_3 r_2 r_3 + p_1 p_2 p_3 r_1 r_2 \nonumber \\
&+& p_1 p_2 p_3 r_1 r_3 + p_1 p_2 p_3 r_2 r_3 + p_1 p_2 r_1 r_2 r_3 \nonumber \\
&+& p_1 p_3 r_1 r_2 r_3 + p_2 p_3 r_1 r_2 r_3 - 2 p_1 p_2 p_3 r_1 r_2 r_3 = 0  \ ,
\label{eq:martini}
\end{eqnarray}
which was also reported recently in \cite{Wu06}. Taking $r_i = 1$, we get the 
result for the critical surface of the general honeycomb lattice \cite{SykesEssam}:
\begin{equation}
1 - p_1 p_2 - p_1 p_3 - p_2 p_3 + p_1 p_2 p_3 =0 \ , \label{eq:honeycomb}
\end{equation}
and taking $p_i = 1$ we get the following formula for the critical
surface of the general triangular lattice \cite{SykesEssam}:
\begin{equation}
1 - r_1 - r_2 - r_3 + r_1 r_2 r_3 =0 \ . \label{eq:triangular}
\end{equation}

\begin{figure}
\begin{center}
\includegraphics[]{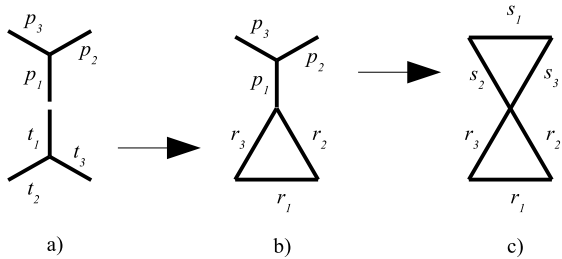}
\caption{The transformation from the (a) double honeycomb, to the (b) martini, to the (c) kagom\'{e} lattice.} \label{fig:hex_martini_kagome}
\end{center}
\end{figure}

For the first approach to the kagom\'e lattice, we start 
with the inhomogeneous double-bond honeycomb lattice, whose unit cell is
shown in Fig.\ \ref{fig:hex_martini_kagome}(a). Replacing the bond with probability $p_i$ in the honeycomb
lattice with a pair of bonds in series with probability $p_i t_i$, we find from (\ref{eq:honeycomb})
that the critical surface is given by
\begin{equation}
1-p_1 p_2 t_1 t_2 - p_2 p_3 t_2 t_3 - p_1 p_3 t_1 t_3 + p_1 p_2 p_3 t_1 t_2 t_3=0 \ . \label{eq:doublehex}
\end{equation}

Now consider the progression shown in Fig.\ \ref{fig:hex_martini_kagome}. Starting with the double honeycomb lattice (a), changing every up star into a triangle gives the martini lattice (b), and changing the down stars gives the kagom\'e lattice (c). The fact that the thresholds of the first two stages of this transformation are now known allows us to make guesses as to the way to reach the third.

Comparing (\ref{eq:doublehex}) with (\ref{eq:martini}),
it can be seen that the transformation
\begin{eqnarray}
t_1 t_2 &\rightarrow& r_3 + r_1 r_2 (1-r_3) \label{eq:first_trans}\\
t_1 t_3 &\rightarrow& r_2 + r_1 r_3 (1-r_2)\\
t_2 t_3 &\rightarrow& r_1 + r_2 r_3 (1-r_1) \label{eq:third_trans}\\
t_1 t_2 t_3 &\rightarrow& r_1 r_2 r_3 + r_1 r_2 (1-r_3) \nonumber\\
&&+ r_2 r_3 (1-r_1)+ r_1 r_3 (1-r_2) \label{eq:fourth_trans}
\end{eqnarray}
effectively turns the double honeycomb critical surface into the martini critical surface.
These substitutions can be interpreted in terms of probabilities of
connections between vertices on a triangle, i.e.,
$t_1 t_2$ is the probability that a particular pair of vertices are connected on the star,
and $r_3 + r_1 r_2 (1-r_3)$ is the probability of the same thing on the triangle. 
The same transformations will also change the critical surface of the honeycomb lattice  (\ref{eq:honeycomb}) into that of the triangular (\ref{eq:triangular}) --- but note that we are not applying the star-triangle transformation here. In fact, these manipulations are largely formal, as the equation (\ref{eq:fourth_trans}) is not implied by (\ref{eq:first_trans}) --- (\ref{eq:third_trans}).
Nevertheless, we conjecture that if we transform the down star the same way,
we will be on the kagom\'e critical surface. Using  (\ref{eq:first_trans})-(\ref{eq:fourth_trans}) with
$t_i$ replaced by $p_i$ and $r_i$ by $s_i$, we find that (\ref{eq:martini}) becomes
\begin{eqnarray}
1 &-& r_1 s_1 - r_2 s_2 - r_3 s_3 - s_1 r_2 r_3- s_2 r_1 r_3 - s_3 r_1 r_2 \nonumber \\
&-& r_1 s_2 s_3 - r_2 s_1 s_3 - r_3 s_1 s_2 + s_1 r_1 r_2 r_3 + s_2 r_1 r_2 r_3 \nonumber \\
&+& s_3 r_1 r_2 r_3 + r_1 r_2 s_1 s_3+ r_1 r_3 s_1 s_2 + r_2 r_3 s_1 s_2  \nonumber \\
&+& r_2 r_3 s_1 s_3 + r_1 r_2 s_2 s_3 + r_2 s_1 s_2 s_3 + r_3 s_1 s_2 s_3 \nonumber \\
&+& r_1 r_3 s_2 s_3 + r_1 s_1 s_2 s_3- r_1 r_2 r_3 s_1 s_3 - r_1 r_2 r_3 s_2 s_3 \nonumber \\
&-& r_1 r_2 r_3 s_1 s_2 - r_1 r_2 s_1 s_2 s_3  - r_1 r_3 s_1 s_2 s_3 \nonumber \\
&-& r_2 r_3 s_1 s_2 s_3 + r_1 r_2 r_3 s_1 s_2 s_3 = 0\ . \label{eq:kagome}
\end{eqnarray}
Setting all probabilities equal gives the condition
\begin{equation}
1 - 3 p^2 - 6 p^3 + 12 p^4 - 6 p^5 + p^6=0 \ , \label{eq:kagomeWu}
\end{equation}
with solution in $[0,1]$ $p_c=0.5244297175...$. This result turns out to be identical to the conjecture
made several years ago by Wu \cite{Wu79} by different means.
Subsequently, this
value was found to be high numerically, but by only $3 \cdot 10^{-5}$ \cite{ZiffSuding97}. Note that (\ref{eq:kagome}) is a plausible form for the kagom\'e threshold: all the bonds are equivalent, setting any one probability to $0$ gives the correct threshold for the $A$ lattice [the lattice that results when $p_1$ is set to $1$ in Fig.\ \ref{fig:martini}(b)], and setting all $p_i=1$ reduces the expression to the triangular critical surface. It is difficult to imagine any other form that satisfies these conditions and remains linear in the probabilities, suggesting that the true general formula for the kagom\'e lattice will not be linear in this way.

\begin{figure}
\begin{center}
\includegraphics[]{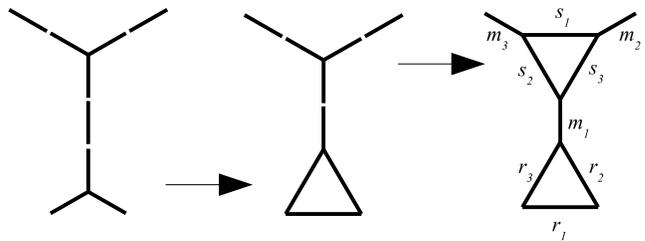}
\caption{Progression from the triple-bond honeycomb to the $(3,12^2)$ lattice.} \label{fig:hex_martini_3_12}
\end{center}
\end{figure}

The same procedure can also be used to find an approximate solution to the $(3,12^2)$ lattice. We start with the {\it triple}-bond honeycomb lattice, and transform the stars into triangles in the same manner as before (Fig.\ \ref{fig:hex_martini_3_12}). There are nine probabilities in this case and the resulting inhomogeneous condition is
\begin{eqnarray}
1&-& m_1 m_2 ( r_3 + r_1 r_2 - r_1 r_2 r_3) (s_3 + s_1 s_2 - s_1 s_2 s_3) \nonumber \\
&-& m_1 m_3 (r_2 + r_1 r_3 - r_1 r_2 r_3) (s_2 + s_1 s_3 - s_1 s_2 s_3) \nonumber \\
&-& m_2 m_3 (r_1 + r_2 r_3 - r_1 r_2 r_3) (s_1 + s_2 s_3 - s_1 s_2 s_3) \nonumber \\
&+& m_1 m_2 m_3 (r_1 r_2 + r_1 r_3 + r_2 r_3 - 2 r_1 r_2 r_3) \nonumber \\
&\times&(s_1 s_2 + s_1 s_3 + s_2 s_3 -
2 s_1 s_2 s_3)=0\ . \label{eq:3_12}
\end{eqnarray}
Setting all $m_i = 1$ gives (\ref{eq:kagome})  (in factored form),  and
setting all $m_i = m$ and $r_i = s_i = r$ gives the equation for
an inhomogeneous $(3,12^2)$ lattice with all triangle bonds having
probability $r$ and all linking bonds having probability $m$:
\begin{equation}
1 - 3 m^2 (r + r^2 - r^3)^2 + m^3 (3 r^2 - 2 r^3)^2 = 0 \ . \label{eq:3_12_reduced}
\end{equation}
Finally, letting $r = m = p$ gives the equation for the homogeneous
$(3,12^2)$ lattice,
\begin{equation}
(1 + p - 2 p^3 + p^4)(1 - p + p^2 + p^3 - 7 p^4 + 4 p^5)=0 \ , 
\end{equation}
with solution on $[0,1]$ $p_c=0.7404233179...$, well within the bounds of (\ref{eq:may_wierman}). According to the numerical analysis of Parviainen \cite{Parviainen},
$p_c(3,12^2)=0.74042195(80)$.
Our result is high by less than two standard deviations. Yet, we can get even better agreement
with both of these results by taking a somewhat different route.

In our second approach, we also compare the critical double honeycomb with the critical martini
lattice, but we consider all bonds equivalent, in which case
the double honeycomb threshold
is $p_0 = \sqrt{1 - 2 \sin \pi/18}$  by (\ref{eq:doublehex}).
Now, consider the martini lattice with $p_1 = p_2 = p_3 = p$,
and $r_1 = r_2 = r_3 = r$.  Equation (\ref{eq:martini}) implies that the critical surface is 
\begin{equation}
1 - 3 p^2 (r + r^2 - r^3) + p^3 (3 r^2 - 2 r^3) = 0 \ .
\label{martini2}
\end{equation}
and taking $p = p_0$, we find that the critical value for $r$ is
\begin{equation}
r = 0.52440876529769\ldots \ .
\label{kagomerz}
\end{equation}
That is, when one star with bond probabilities $p_0$ is 
replaced by a triangle with probabilities $r$, the system remains at a critical point
(even though local correlations will necessarily be different because this is not
a fixed point of the star-triangle transformation).  If we conjecture that the system
still remains at a critical point when we make the same replacement for the other
triangle, then (\ref{kagomerz})
is an estimate for the $p_c$ of the
kagom\'e lattice. In fact, (\ref{kagomerz}) is very close to the numerical 
result, $p_c = 0.5244053(3)$ \cite{ZiffSuding97}, although outside the 
given error bars.

It turns out that (\ref{kagomerz}) is numerically identical to the value conjectured
by Hori and Kitahara, which however is only available as a conference abstract \cite{HoriKitahara}, without a derivation. Evidently,
we have effectively duplicated the derivation of these authors.
\begin{figure}
\begin{center}
\includegraphics[]{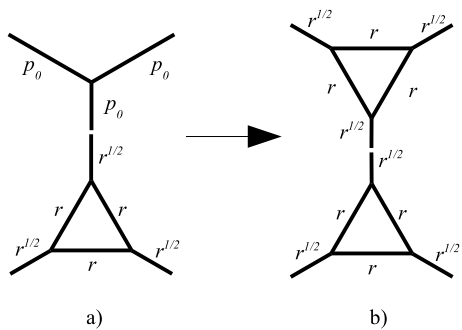}
\caption{The substitution of probabilities for the second $(3,12^2)$ lattice threshold estimate.} \label{fig:hex_to_3_12}
\end{center}
\end{figure}
However, we can go further and use our argument to estimate the threshold for the $(3,12^2)$ lattice.
Again we start off with the double honeycomb lattice at the uniform threshold
of $p_0$, and compare to a critical martini lattice with $p = p_0\,\sqrt{r}$ [Fig.\ \ref{fig:hex_to_3_12}(a)]. The argument works as in the kagom\'e case, with the transformation to the $(3,12^2)$ lattice shown in Fig.\ \ref{fig:hex_to_3_12}(b). The solution to (\ref{martini2}) yields
\begin{equation}
r = 0.74042117858374\ldots  \ . \label{eq:3_12rz}
\end{equation}
This result is within
the error bars of \cite{Parviainen} and
falls within the rigorous bounds of \cite{Wierman03},
which raises
the possibility that the result is exact. Clearly, more precise numerical work
for both lattices is called for.

We can generalize our argument above for the inhomogeneous
$(3,12^2)$ lattice with two probabilities $m$ and $r$.  The
critical surface is determined by (\ref{martini2}) with $p = p_0 \sqrt{m}$.
When $m = 1$, this gives the kagom\'e estimate (\ref{kagomerz}),
when $m = r$ it gives the homogeneous estimate (\ref{eq:3_12rz}), and
when $r =1$ it gives the exact honeycomb result $m = p_0^2$.
The formula (\ref{martini2}) (with $p = p_0 \sqrt{m}$) can be compared
with (\ref{eq:3_12_reduced}), which
though mathematically quite different, gives very similar numerical
solutions.  Finally, we note one last relation: if we require
that the second terms of the two estimates (\ref{eq:3_12_reduced}) and (\ref{martini2})
(which represents two-point correlations) be the same, we get
the simple condition
\begin{equation}
p_0^2 / m = r + r^2 - r^3
\end{equation}
which turns out to be identical to Tsallis' conjecture for this system.
As mentioned above, however, the predictions of this formula are much farther
from the numerical measurements than the predictions of (\ref{eq:3_12_reduced})
and (\ref{martini2}). 

In conclusion, we have shown that
the results for the martini and honeycomb lattices can be
used to make precise estimates 
of bond percolation on the kagom\'e and $(3,12^2)$ lattices, both long-standing problems in percolation theory. For the kagome lattice, we have reproduced the conjectures of both Wu and of Hori and Kitahara, while for the $(3,12^2)$ lattice we have
apparently very precise estimates. Perhaps these methods can point the way to
finding rigorous thresholds for these lattices, and analyzing other unsolved lattices in
percolation.

The authors thank W. May
and J. Wierman for providing a copy of their preprint prior to
publication. R.Z. acknowledges support from NSF grant DMS0244419.

\bibliography{ziff-scullard}

\end{document}